\documentclass[10pt,english]{extarticle}
\usepackage[utf8]{inputenc}
\usepackage{babel}
\usepackage{amsmath}
\usepackage{amsthm}
\usepackage{amssymb}
\usepackage{graphicx}
\usepackage[unicode=true,pdfusetitle,
 bookmarks=true,bookmarksnumbered=false,bookmarksopen=false,
 breaklinks=false,pdfborder={0 0 1},backref=false,colorlinks=false]
 {hyperref}

\makeatletter
\numberwithin{equation}{section}
\numberwithin{figure}{section}
\theoremstyle{plain}
\newtheorem{thm}{\protect\theoremname}
\theoremstyle{definition}
\newtheorem{example}[thm]{\protect\examplename}
\newcommand{\lyxaddress}[1]{
	\par {\raggedright #1
	\vspace{1.4em}
	\noindent\par}
}

\makeatother

\usepackage{listings}
\providecommand{\examplename}{Example}
\providecommand{\theoremname}{Theorem}

\begin{document}
\title{Differentiable programming for particle physics simulations}
\author{Roland Grinis\thanks{Moscow Institute of Physics and Technology}
\thanks{GrinisRIT ltd.} }
\maketitle
\begin{abstract}
We describe how to apply adjoint sensitivity methods to backward Monte-Carlo
schemes arising from simulations of particles passing through matter.
Relying on this, we demonstrate derivative based techniques for solving
inverse problems for such systems without approximations to underlying
transport dynamics. We are implementing those algorithms for various
scenarios within a general purpose differentiable programming \textsf{C++17}
library \textsf{NOA} (\textsf{github.com/grinisrit/noa}).
\end{abstract}

\section{Introduction}

In this paper, we explore the challenges and opportunities that arise
in integrating \emph{differentiable programming} (DP) with simulations
in particle physics.

In our context, we will broadly refer to DP as a program for which
some of the inputs could be given the notion of a variable, and the
output of that program could be differentiated with respect to them.

Most common examples include the widely used \emph{deep learning}
(DL) models created over the powerful \emph{automatic differentiation}
(AD) engines such as \textsf{TensorFlow} and \textsf{PyTorch}. Since
their initial release, those \emph{machine learning} (ML) frameworks
grew up into fully-fledged DP libraries capable of tackling a more
diversified set of tasks.

Recently, a very fruitful interaction between DP as we know it in
ML and numerical solutions to differential equations started to gather
pace with the work of R. Chen et al. \cite{chen}. A whole new area
tagged now-days \emph{Neural Differential Equations }arose in scientific
ML. 

On one hand, using ML we obtain a more flexible framework with a wealth
of new tools to tackle a variety of \emph{inverse problems} in mathematical
modeling. On the other hand, many techniques in the latter such as
the \emph{adjoint sensitivity methods} give rise to new powerful algorithms
for AD.

A few implementations are now available:
\begin{itemize}
\item \textsf{torchdiffeq} is the initial \textsf{python} package developed
by \cite{chen} providing ODE solvers that not only integrate with
\textsf{PyTorch} DL models, but also use those to describe the dynamics.
\item \textsf{torchsde} builds off from \textsf{torchdiffeq} and provides
the same functionality for SDEs, as well as $\mathcal{O}(1)$-memory
gradient computation algorithms, see \cite{chen2}.
\item \textsf{diffeqflux }is a \textsf{Julia }package developed by C. Rackauckas
et al. \cite{chris} and relies on a rich scientific ML ecosystem
treating many different types of equations including PDEs.
\end{itemize}
Unsurprisingly, one can also find roots of this story in \emph{computational
finance}, see for example the work of M. Giles et al. \cite{mike}.
An AD algorithm is presented there for computing the risk sensitivities
for a portfolio of options priced through Monte-Carlo simulation.
That set-up is close to our case of interest and therefore represents
a great source of inspiration for us.

In fact, for particle physics simulations a similar picture is left
almost unexplored so far. The dynamics are richer than the ones considered
before, but we also have more tools at our disposal such as the \emph{Backward
Monte-Carlo} (BMC) techniques. We make use of the latter to adapt
the adjoint sensitivity methods to the transport of particles through
matter simulations.

Ultimately, we obtain a novel methodology for \emph{image reconstruction}
problems when the absorption mechanism is non-linear. In future work,
we will demonstrate this approach in the specific case of \emph{muography}.
We are releasing our implementations within the open source library
\textsf{NOA} \cite{noa}.

\section{Backward Monte-Carlo}

This Monte-Carlo technique seeks to reverse the simulation flow from
a given final state up to a distribution of initial states. 

Several implementations have been considered, including space radiation
problems by L. Desorgher et al. \cite{geant} and more recently muon
transport by V. Niess et al. \cite{valentin}. The latter is backed
by a dedicated \textsf{C99} library \textsf{PUMAS}. We shall recall
here the main set-up and refer the reader to \cite{valentin} for
a more in depth introduction. 

In general, one can represent the transport of particles through matter
as a stochastic flow $\varphi$ on a state space $\textbf{S}$ that
might include observables such as position coordinates, momentum direction
and kinetic energy.

The whole purpose of the simulation is to compute the stationary flux
$\phi$ of particles in some given region of interest in the state
space $\textbf{S}$. For example, in the special case of muography,
that might be a single space point with a fibre of angles for momentum
directions representing the readings on a scintillation detector.
As an aside, we note that the latter is unable to determine the kinetic
energy of particles as of now.

Considering the transition distribution $\tau$ induced by the flow
$\varphi$, one can compute the flux $\phi$ via the convolution:

\begin{equation} \phi(\textbf{s}_f) = \int \tau(\textbf{s}_f; \textbf{s}_i) \phi(\textbf{s}_i)d\textbf{s}_i \end{equation}at
a given final state $\textbf{s}_f \in \textbf{S}$. The BMC sampling
schemes aim to evaluate the above integral. The stochastic flow $\varphi$
is evolved far enough as to reach a region of the state space $\textbf{S}$
where the flux $\phi$ is known. 

If the map $\varphi$ is invertible, one can estimate:

\begin{equation} \phi(\textbf{s}_f) \simeq \frac{1}{N} \sum_{k=1}^N \omega_{k} \phi(\textbf{s}_{i,k}) \end{equation}
where $\textbf{s}_{i,k} = \varphi^{-1}(\textbf{s}_f; x_k)$ for some
independent random variate $x_k$ accounting for the stochasticity
of $\varphi$ and: 

\begin{equation} \omega_{k} = \det(\nabla_{\textbf{s}_f}\varphi^{-1})|_{x_k}. \end{equation}

In practice, the flow $\varphi$ is broken down into a sequence of
$n$ steps:

\begin{equation} \textbf{s}_{0,k} = \varphi_1^{-1} \circ \varphi_2^{-1} \circ \dots \circ \varphi_n^{-1} (\textbf{s}_{n,k}; x_{n,k}) \end{equation}
with $\textbf{s}_{n,k} = \textbf{s}_f$, and so we get: 

\begin{equation} \omega_{k} = \prod_{j=1}^{n}\det(\nabla_{\textbf{s}_{j,k}}\varphi_j^{-1})|_{x_{j,k}}. \end{equation}

Unfortunately, the simulation flow is not always invertible and one
has to rely on \emph{biasing techniques}. In such situation, we need
to construct a regularised version $\varphi_{b}$ of the flow, together
with its transition density $\tau_b$. Then, provided the Radon--Nikodym
derivative of $\tau$ exists w.r.t. $\tau_b$ we can set: 

\begin{equation} \omega_{k} = \frac{\tau(\textbf{s}_f; \textbf{s}_{i,k})}{\tau_b(\textbf{s}_f; \textbf{s}_{i,k})} \det(\nabla_{\textbf{s}_f}\varphi_b^{-1})|_{x_k} \end{equation}
where the initial state is obtained as $\textbf{s}_{i,k} = \varphi_b^{-1}(\textbf{s}_f; x_k)$.

Of special interest to us is the case of mixture distributions: 

\begin{equation} \tau(\textbf{s}_f; \textbf{s}_i) = \sum_{\ell=1}^{m} p_{\ell}(\textbf{s}_i) \tau_{\ell}(\textbf{s}_f; \textbf{s}_i) \end{equation}
for a partition of unity of the state space $\sum_{{\ell}=1}^{m} p_{\ell} \equiv 1$.
For example in the \emph{mixture of materials} case, at each point
$\textbf{s}$ we choose whether to interact with material $\ell$
with probability $p_{\ell}(\textbf{s})$.

One proceeds by constructing an \emph{apriori} partition of unity
$\sum_{\ell=1}^{m} p_{\ell, b} \equiv 1$. It is used to choose the
component $\ell_0$ to evolve the flow backwards, giving: 

\begin{equation}
\label{bmcMain}
\omega_{k} = \frac{p_{\ell_0}(\textbf{s}_{i,k})}{p_{\ell_0,b}(\textbf{s}_f)} \det(\nabla_{\textbf{s}_f}\varphi_{\ell_0}^{-1})|_{x_{\ell_0,k}} \end{equation}with $\textbf{s}_{i,k} = \varphi_{\ell_0}^{-1}(\textbf{s}_f; x_k)$.

Of course, if the map $\varphi_{\ell_0}$ is not invertible itself,
one needs to construct a regularised map $\varphi_{\ell_0, b}$, compute
 $\textbf{s}_{i,k} = \varphi_{\ell_0,b}^{-1}(\textbf{s}_f; x_k)$
and set:

\begin{equation}
\omega_{k} = \frac{p_{\ell_0}(\textbf{s}_{i,k})}{p_{\ell_0,b}(\textbf{s}_f)}\frac{\tau_{\ell_0}(\textbf{s}_f; \textbf{s}_{i,k})}{\tau_{\ell_0,b}(\textbf{s}_f; \textbf{s}_{i,k})} \det(\nabla_{\textbf{s}_f}\varphi_{\ell_0,b}^{-1})|_{x_k}.
\end{equation}

Equation \ref{bmcMain} is our starting point for integrating differentiable
programming with BMC. Following \cite{mike}, the idea is to commute
differentiation with MC sampling by performing a change of measure
which is independent of the variables. In our case, if we allow the
partition of unity to depend on some variable $\vartheta$: 

\begin{equation} 
\omega_{k}(\vartheta) = \frac{p_{\ell_0}(\textbf{s}_i, \vartheta)}{p_{\ell_0,b}(\textbf{s}_f)} \det(\nabla_{\textbf{s}_f}\varphi_{\ell_0}^{-1})|_{x_{\ell_0,k}} 
\end{equation} then the required change of measure is simply provided by the biasing
scheme we are already using to invert the flow. We have been careful
to choose the biasing partition of unity $p_{\ell,b}$ independent
of the variable $\vartheta$. This type of regularisation can be achieved
by picking analytical functions, such as Gaussians which are never
vanishing and fast decaying. 

Naturally, $\vartheta$ represents the target to estimate for image
reconstruction tasks when $p_{\ell}$ describes materials mixture.

\section{Differentiable Programming}

\begin{figure}
\begin{centering}
\includegraphics[width=12cm]{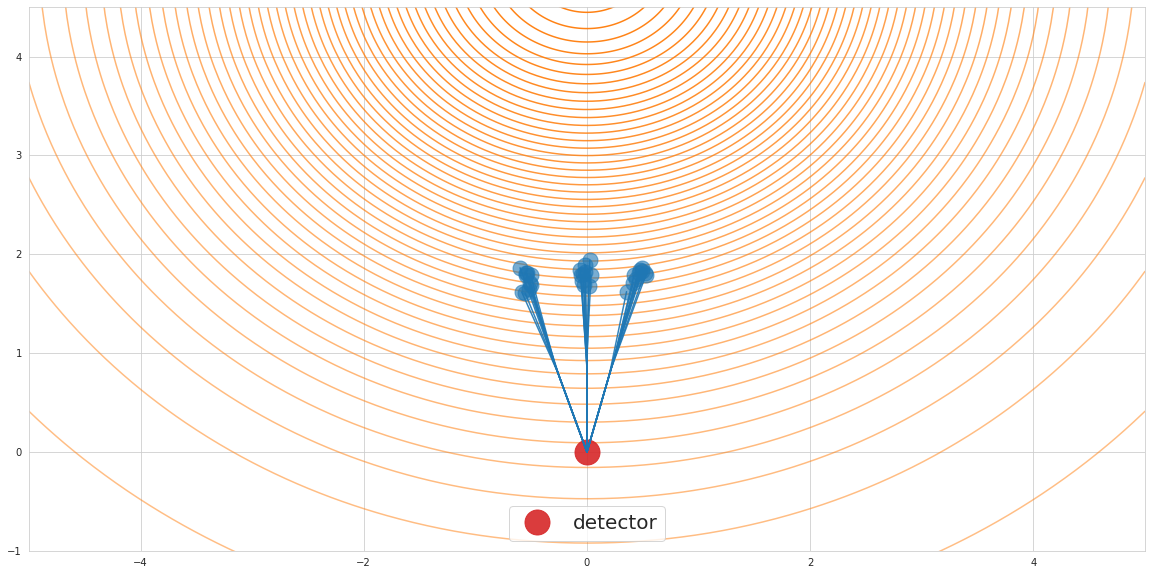}
\par\end{centering}
\caption{\label{fig:Two-step-BMC} The contours correspond to level sets of
the materials mixture with $\vartheta_{\mu} = \left(0, 5 \right)$
and $\vartheta_{\sigma}= \sqrt{10}$. In blue we show the BMC simulated
trajectories. For the sake of simplicity, we assume the known particle
flux is constant equals to one and is reached after two steps. The
measurement angles on the detector have been arbitrarily chosen at
$\left(-\frac{\pi}{5},0,\frac{4\pi}{25}\right)$.}
\end{figure}

In this section we will present a toy example, but one which will
illustrate the key aspects of the algorithm. We invite the reader
to look at the accompanying notebook \textsf{differentiable\_programming\_pms.ipynb}
in \textsf{NOA} \cite{noa} to reproduce the results stated here.
\begin{example}
\label{exa:basic} We get ourselves into a two dimensional space with
the \emph{detector} placed at the origin, see figure \ref{fig:Two-step-BMC}.
We take a partition of unity parameterised by a single Gaussian kernel: 

\begin{equation}  
1 = \exp\left(- \frac{\|\textbf{s} - \vartheta_{\mu}\|^2}{\vartheta_{\sigma}^2} \right) + \left[ 1 - \exp\left(- \frac{\|\textbf{s} - \vartheta_{\mu}\|^2}{\vartheta_{\sigma}^2} \right) \right]
\end{equation}for a position only state $\mathbf{s}\in\mathbb{R}^{2}$ and parameters
$\vartheta_{\mu} \in \mathbb{R}^2$, $\vartheta_{\sigma} > 0$ to which
we typically refer as simply $\vartheta = \left[ \vartheta_{\mu} || \vartheta_{\sigma}\right]$.

Let us suppose that the material corresponding to the Gaussian is
easy to penetrate and has an attenuation coefficient of 10\%. But
for the background we set the attenuation to 99\%. We have an extreme
contrast between the two and we want to localise the first one given
measurements at different angles on our detector. 

A naive implementation of the BMC scheme to compute the flux in this
configuration can be found in the appendix, routine \textsf{backward\_mc}
\ref{exa:AD-implementation}.
\end{example}

If the parameters $\vartheta$ are fixed, the distribution of the
flux approaches a normal one with mean estimated by: 

\begin{equation} 
\hat\phi(\textbf{s}_f; \vartheta) = \frac{1}{N} \sum_{k=1}^N \omega_{k}(\vartheta) \cdot \phi(\textbf{s}_{i,k}) 
\end{equation}and variance $\sigma_{N}^2=\mathcal{O}(1/N)$ as the number of particles
$N \rightarrow \infty$, which we treat as fixed for all purposes. 

Let us postulate that the \emph{random error} in our model follows
a normal distribution $N(0, \sigma_{M}^2)$. Taking a Bayesian approach,
we give a normal distribution $\vartheta \sim N(\mu_\vartheta, \sigma_{\vartheta}^2)$
to the prior as well. Hence, the log-probability for observing a flux
$\phi(\textbf{s}_f)$ on the detector evaluates to: 

\begin{equation} 
\mathcal{L}(\vartheta) = - \frac{\|\phi(\textbf{s}_f) - \hat\phi(\textbf{s}_f; \vartheta)\|^2}{\sigma_{M}^2} - \frac{\|\vartheta- \mu_\vartheta\|^2}{\sigma_{\vartheta}^2} 
\end{equation} up to a constant.

\begin{figure}
\begin{centering}
\includegraphics[width=12cm]{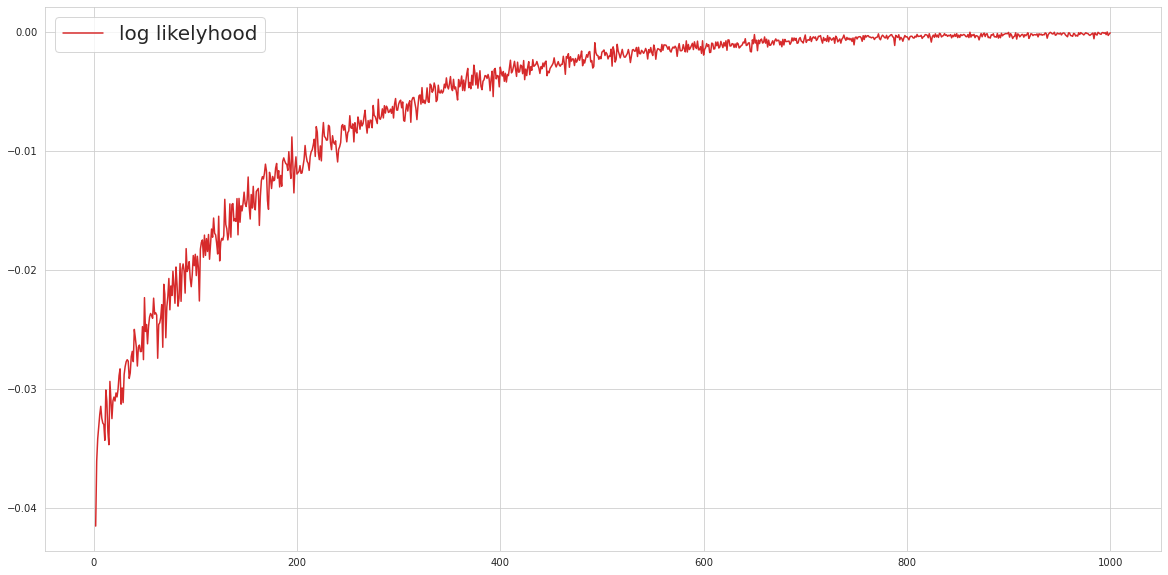}
\par\end{centering}
\caption{\label{fig:SGD-convergence-over}SGD convergence over 1000 steps with
learning rate 0.05 }

\end{figure}

\begin{example}
\label{exa:SGD}We note that the routine \textsf{backward\_mc} \ref{exa:AD-implementation}
is a differentiable program built on top of \textsf{LibTorch}'s \textsf{Autograd}
library. It can be differentiated using \textsf{torch::autograd::grad},
which enables us to run a \emph{Stochastic Gradient Descent} (SGD)
optimisation to solve the inverse problem for $\vartheta$. This can
be presented as a \emph{maximum likelihood estimation} letting $\sigma_{\vartheta} \rightarrow \infty$. 

Recalling figure \ref{fig:Two-step-BMC}, let us set $\bar\vartheta_{\mu} = \left(-1, 5 \right)$
and $\bar\vartheta_{\sigma}= \sqrt{10}$ as our \emph{true values}
and compute the \emph{observed flux} for them. Taking out one measurement,
at angle 0 say for validation later, we keep in the other two $\left(-\frac{\pi}{5},\frac{4\pi}{25}\right)$
for SGD.

Running the classical SGD update:
\begin{equation}
\vartheta\leftarrow\vartheta+\eta\nabla_{\vartheta}\mathcal{L}
\end{equation}
starting from $\vartheta_{\mu} = \left(0, 5 \right)$ we obtain the
convergence graph in figure \ref{fig:SGD-convergence-over}. Taking
the mean over the last 200 values we obtain the \emph{optimal} parameters
$\hat\vartheta_{\mu} = \left(-1.0033,  4.9891 \right)$ and $\hat\vartheta_{\sigma}^2= 9.9611$
which are in good agreement with the true values $\bar\vartheta$
as expected. In future works, we will compare this approach against
other reconstruction algorithms, with genuine dynamics from particle
physics.
\end{example}

To close this section, we would like to demonstrate how one might
explore further the fact that \textsf{backward\_mc} \ref{exa:AD-implementation}
is a differentiable program. 

\begin{figure}

\begin{centering}
\includegraphics[width=12cm]{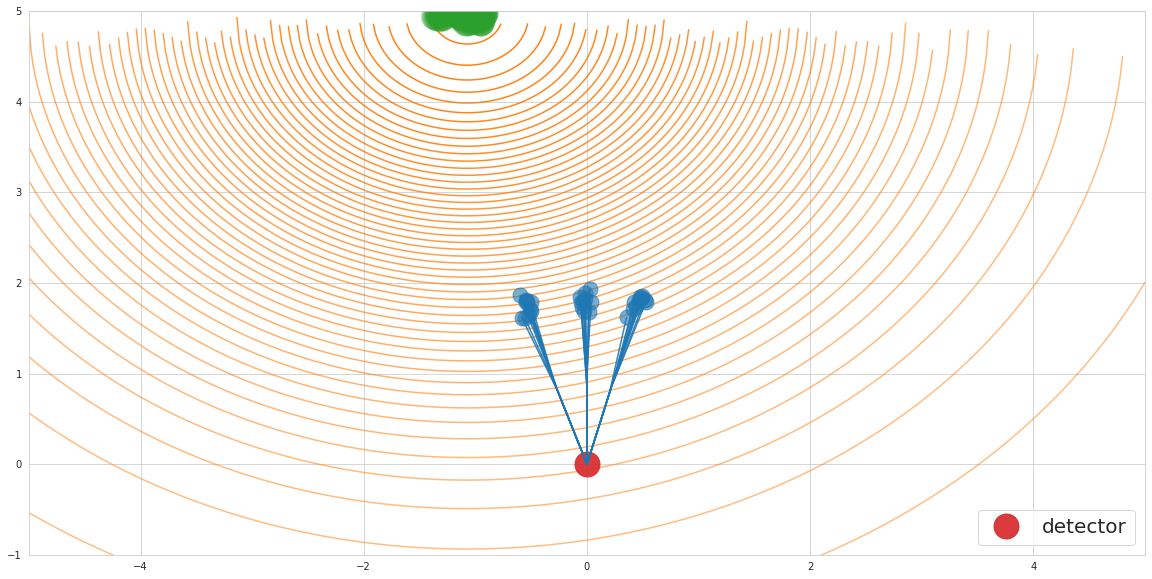}\caption{\label{fig:hmc}Optimal parameters and Bayesian Posterior sample}
\par\end{centering}
\end{figure}

\begin{example}
In example \ref{exa:SGD}, we have left out from the optimisation
the measurement at angle 0. We will use it to look at the stability
of the optimal solution we found with Bayesian inference. For higher
dimensional distributions exhibiting rough curvature standard \emph{Markov
Chain Monte-Carlo} schemes are not suitable. On has to rely on \emph{Hamiltonian
Monte-Carlo} (HMC) which uses Hamiltonian dynamics to build the MC
chain.

The library \textsf{NOA} \cite{noa} implements Riemannian HMC with
an explicit symplectic integrator as in \cite{rmhmc}, \cite{hamiltorch}.
The Hamiltonian $H$ uses the log-probability density $\mathcal{L}(\vartheta)$
as potential, where $\vartheta \in \mathbb{R}^d$ denote the parameters
which we augment with momentum coordinate $\chi \in \mathbb{R}^d$: 

\begin{equation} 
H(\vartheta,\chi) = \frac{1}{2} \chi^t M(\vartheta)^{-1}\chi + \frac{1}{2} \log \det(M(\vartheta)) - \mathcal{L}(\vartheta). 
\end{equation} 

Following M. Betancourt \cite{betan}, the local metric is obtained
applying a regularisation procedure in the form of the \emph{softabs
map}: 

\begin{equation} 
M = Q \cdot\lambda_d \coth (\alpha\lambda_d ) \cdot Q^t 
\end{equation} where $Q(\vartheta), \lambda_d(\vartheta)$ stand for the \emph{eigendecomposition}
of $\nabla^2\mathcal{L}(\vartheta)$, and typically we set $\alpha=10^6$
for the \emph{softabs constant}. Therefore, we are indirectly using
the \emph{3rd order derivative} of our model for the flux. 

This Hamiltonian is non-separable, and to avoid using an implicit
integrator which is computationally heavy, one can rely on the explicit
algorithm from M. Tao \cite{tao}. The phase space $\vartheta, \chi$
is augmented by $\vartheta^*, \chi^* \in \mathbb{R}^d$ and we solve
the corresponding separable Hamiltonian dynamics: 

\begin{equation} 
H^*(\vartheta,\chi, \vartheta^*, \chi^*) = H(\vartheta,\chi^*) + H(\vartheta^*,\chi) + \frac{1}{2}\Omega \cdot(\|\vartheta-\vartheta^*\|^2_2 + \|\chi-\chi^*\|^2_2) 
\end{equation}where the constant $\Omega$ needs to be tuned.

We sample 300 points via the HMC scheme for $\mathcal{L}(\vartheta)$
defined using the reading at angle 0 only but with the prior mean
given by the optimal solution found in example \ref{exa:SGD}. The
result is illustrated in figure \ref{fig:hmc}. We reconfirm our result
with a posterior sample mean $\hat\vartheta_{\mu} = \left(-1.0722,  4.9530 \right) \pm \left(0.1343, 0.0410 \right)$.

One important direction for improvement here is to introduce the appropriate
form of \emph{Stochastic Gradient Riemannian HMC} since the log-probability
density is evaluated through backward MC.
\end{example}

\section{Adjoint sensitivity methods}

There is a challenge with the above approach differentiating through
the MC simulation using AD. Our algorithm is not constant in the number
of steps $n$ for the discretisation of the transport. This issue
can be addressed by adjoint sensitivity methods \cite{pontrjagin}. 

Let us recall how this technique works for an ODE: 

\begin{equation} 
\frac{dx}{dt} = f(t, x(t), \vartheta). 
\end{equation}

Imagine that we want to compute the gradient of some scalar function: 

\begin{equation} 
\mathcal{L}(x(t_1)) = \mathcal{L} \left( x(t_0) + \int_{t_0}^{t_1}f(t, x(t), \vartheta)dt\right) 
\end{equation}with respect to the parameters $\vartheta$. 

An efficient way to tackle this task is to introduce the adjoint: 

\begin{equation} a(t) = \nabla_{x(t)}\mathcal{L} \end{equation} which
satisfies the adjoint ODE: 

\begin{equation} \frac{da}{dt} = - a(t) \cdot \nabla_x f(t, x(t), \vartheta). \end{equation}

The desired gradient is given then by a backward in time integral: 

\begin{equation} 
\nabla_{\vartheta}\mathcal{L} = - \int_{t_1}^{t_0} a(t) \cdot \nabla_{\vartheta} f(t, x(t), \theta)dt. 
\end{equation}

In our set-up, we are simulating the evolution back from the final
state straightaway. Therefore, we can easily adapt the adjoint sensitivity
method to the BMC scheme. 

In doing so, we recall that the weight is computed iteratively:

\begin{equation} 
\omega_{j+1}(\vartheta) = \frac{p_{\ell_0}(\textbf{s}_{j+1}, \vartheta)}{p_{\ell_0,b}(\textbf{s}_j)} \det(\nabla_{\textbf{s}_j}\varphi_{j,\ell_0}^{-1})|_{x_{j,\ell_0}} \cdot \omega_{j}(\vartheta) 
\end{equation} for $j = 0 .. n-1$, with $\omega_0 = 1 $ and $\omega_n$ giving
the final weight. 

The \emph{discrete adjoint sensitivity algorithm} yielding $\mathcal{O}(1)$-memory
derivative computations is simply obtained by differentiating through
the above recursion:

\begin{equation}  
\nabla_{\vartheta}^m \omega_{j+1}(\vartheta) = \det(\nabla_{\textbf{s}_j}\varphi_{j,\ell_0}^{-1})|_{x_{j,\ell_0}} \sum_{k=0}^{m} {m \choose k} \frac{\nabla_{\vartheta}^k p_{\ell_0}(\textbf{s}_{j+1}, \vartheta)}{p_{\ell_0,b}(\textbf{s}_j)} \cdot \nabla_{\vartheta}^{m-k}\omega_{j}(\vartheta).  \end{equation}

To evaluate $\nabla_{\vartheta}^k p_{\ell_0}$ one will still rely
on the AD engine. However, the computational graph needed for reverse-mode
differentiation can be freed after each step and executed in parallel
for each particle. 

The routine \textsf{backward\_mc\_grad} \ref{exa:ADS} in the appendix
provides an implementation with first order derivative for the BMC
scheme discussed in example \ref{exa:basic}.

\section{Conclusion}

In this paper, we have demonstrated how to efficiently integrate automatic
differentiation and adjoint sensitivity methods with\emph{ }Backward
Monte-Carlo schemes arising in the passage of particles through matter
simulations. 

We believe that this builds a whole new bridge between scientific
machine learning and inverse problems arising in particle physics.
In future, we hope to prove the success of this technique in a variety
of image reconstruction problems with non-linear dynamics, starting
with muography.

\subsection*{Acknowledgments}

We would like to thank the \textsf{MIPT-NPM} lab and Alexander Nozik
in particular for very fruitful discussions that have led to this
work. We are very grateful to \textsf{GrinisRIT} for the support.
This work has been initially presented in June 2021 at the QUARKS
online workshops 2021 - \textquotedblleft \emph{Advanced Computing
in Particle Physics}\textquotedblright .

\lyxaddress{\emph{Email}: \textsf{roland.grinis@grinisrit.com}}

\lyxaddress{Moscow Institute of Physics and Technology, Institutsky lane 9, Dolgoprudny,
Moscow region, Russia, 141700}

\section*{Appendix: Code examples}

We have collected here the two BMC implementations for our basic example.
You can reproduce all the calculations in this paper from the notebook
\textsf{differentiable\_programming\_pms.ipynb} available in \textsf{NOA}
\cite{noa}. 

For the specific code snippets here the only dependency is \textsf{LibTorch:}

\begin{lstlisting}[language={[GNU]C++},basicstyle={\footnotesize\sffamily},tabsize=2]
#include <torch/torch.h>
\end{lstlisting}

The following routine will be used throughout and provides the rotations
by a tensor \textsf{angles} for multiple scattering:

\begin{lstlisting}[language={[GNU]C++},basicstyle={\footnotesize\sffamily},tabsize=4]
inline torch::Tensor rot(const torch::Tensor &angles) 
{
	 const auto n = angles.numel();
	 const auto c = torch::cos(angles);
	 const auto s = torch::sin(angles);
	 return torch::stack({c, -s, s, c}).t().view({n, 2, 2});
}
\end{lstlisting}

Given the set-up in example \ref{exa:basic} we define:

\begin{lstlisting}[language={[GNU]C++},basicstyle={\footnotesize\sffamily},tabsize=4]
const auto detector = torch::zeros(2);     
const auto materialA = 0.9f;     
const auto materialB = 0.01f;

inline const auto PI = 2.f * torch::acos(torch::tensor(0.f));

inline torch::Tensor mix_density(
	const torch::Tensor &states, 
	const torch::Tensor &vartheta) 
{     
	return torch::exp(-(states - vartheta.slice(0, 0, 2))
		.pow(2).sum(-1) / vartheta[2].pow(2)); 
}
\end{lstlisting}

\begin{example}
\label{exa:AD-implementation}This implementation relies completely
on the AD engine for tensors. The whole trajectory is kept in memory
to perform reverse-mode differentiation. 

The routine accepts a tensor \textsf{theta }representing the angles
for the readings on the detector, the tensor \textsf{node }encoding
the mixture of the materials which is essentially our variable, and
the number of particles \textsf{npar. }

It outputs the simulated flux on the detector corresponding to \textsf{theta}: 

\begin{lstlisting}[language={[GNU]C++},basicstyle={\footnotesize\sffamily},breaklines=true,showstringspaces=false,tabsize=4]
inline torch::Tensor backward_mc(
	const torch::Tensor &theta,         
	const torch::Tensor &node,         
	const int npar) 
{
	const auto length1 = 1.f - 0.2f * torch::rand(npar);     
	const auto rot1 = rot(theta);

	auto step1 = torch::stack({torch::zeros(npar), length1}).t();
	step1 = rot1.matmul(step1.view({npar, 2, 1})).view({npar, 2});
	const auto state1 = detector + step1;
	
	auto biasing = torch::randint(0, 2, {npar});    
	auto density = mix_density(state1, node);   
	auto weights =            
		torch::where(biasing > 0,                         
					(density / 0.5) * materialA,
					((1 - density) / 0.5) * materialB) * 
							torch::exp(-0.1f * length1);
	
	const auto length2 = 1.f - 0.2f * torch::rand(npar);     
	const auto rot2 = rot(0.05f * PI * (torch::rand(npar) - 0.5f));    
	auto step2 = 
		length2.view({npar, 1}) * step1 / length1.view({npar, 1});    
	
	step2 = rot2.matmul(step2.view({npar, 2, 1})).view({npar, 2});     
	const auto state2 = state1 + step2;
	
	biasing = torch::randint(0, 2, {npar});     
	density = mix_density(state2, node);     
	weights *=            
		torch::where(biasing > 0,                          
					(density / 0.5) * materialA,                       
					((1 - density) / 0.5) * materialB) * 
						torch::exp(-0.1f * length2);

	// assuming the flux is known equal to one at state2
	return weights;
}
\end{lstlisting}
\end{example}

\newpage{}
\begin{example}
\label{exa:ADS} This routine adopts the adjoint sensitivity algorithm
to earlier example \ref{exa:AD-implementation}. It outputs the value
of the flux and the first order derivative w.r.t. the tensor \textsf{node:}
\begin{lstlisting}[language={[GNU]C++},basicstyle={\footnotesize\sffamily},tabsize=4]
inline std::tuple<torch::Tensor, torch::Tensor> backward_mc_grad(         
	const torch::Tensor &theta,         
	const torch::Tensor &node) 
{     
	const auto npar = 1; //work with single particle         
	auto bmc_grad = torch::zeros_like(node);

	const auto length1 = 1.f - 0.2f * torch::rand(npar);    
	const auto rot1 = rot(theta);     
	auto step1 = torch::stack({torch::zeros(npar), length1}).t();     
	step1 = rot1.matmul(step1.view({npar, 2, 1})).view({npar, 2});     
	const auto state1 = detector + step1;

    auto biasing = torch::randint(0, 2, {npar});
	auto node_leaf = node.detach().requires_grad_();     
	auto density = mix_density(state1, node_leaf);     
	auto weights_leaf = torch::where(biasing > 0,                          
		(density / 0.5) * materialA,                          
		((1 - density) / 0.5) * materialB) * torch::exp(-0.01f * length1);  
	
	bmc_grad += torch::autograd::grad({weights_leaf}, {node_leaf})[0];     
	auto weights = weights_leaf.detach();

	const auto length2 = 1.f - 0.2f * torch::rand(npar);     
	const auto rot2 = rot(0.05f * PI * (torch::rand(npar) - 0.5f));     
	auto step2 = length2.view({npar, 1}) * step1 / length1.view({npar, 1});     
	step2 = rot2.matmul(step2.view({npar, 2, 1})).view({npar, 2});     
	const auto state2 = state1 + step2;
	
	biasing = torch::randint(0, 2, {npar});
	node_leaf = node.detach().requires_grad_();     
	density = mix_density(state2, node_leaf);     
	weights_leaf = torch::where(biasing > 0,                          
		(density / 0.5) * materialA,                          
		((1 - density) / 0.5) * materialB) * torch::exp(-0.01f * length2); 
   
	const auto weight2 = weights_leaf.detach();     
	bmc_grad = weights * torch::autograd::grad({weights_leaf}, {node_leaf})[0] 
		+ weight2 * bmc_grad;     
	weights *= weight2;
	
	// assuming the flux is known equal to one at state2     
	return std::make_tuple(weights, bmc_grad); 
}
\end{lstlisting}
\end{example}


\begin{thebibliography}{9}

\bibitem{noa} Differentiable Programming for Optimisation Algorithms over LibTorch. \textsf{https://github.com/grinisrit/noa}

\bibitem{betan} M. Betancourt. A general metric for Riemannian manifold Hamiltonian Monte Carlo. \textit{International Conference on Geometric Science of Information}, Springer, 327-334, 2013.

\bibitem{mike} L. Capriotti and M. B. Giles. Algorithmic differentiation: Adjoint greeks made easy. \textit{SSRN Electronic Journal}, 2011.

\bibitem{chen} R. T. Q. Chen, Y. Rubanova, J. Bettencourt, and D. K. Duvenaud. Neural ordinary differential equations. \textit{Advances in neural information processing systems}, 6571-6583, 2018.

\bibitem{hamiltorch} A. Cobb, A. Baydin, A. Markham, and S. Roberts. Introducing an explicit symplectic integration scheme for Riemannian manifold Hamiltonian Monte-Carlo. \textit{preprint arXiv:1910.06243}, 2019.

\bibitem{geant} L. Desorgher, F. Lei, and G. Santin. Implementation of the reverse/adjoint Monte Carlo method into Geant4. \textit{Nucl. Instrum. Meth.}, A621:247-257, 2010.

\bibitem{rmhmc} M. Girolami and B. Calderhead. Riemann manifold Langevin and Hamiltonian Monte Carlo methods. \textit{Journal of the Royal Statistical Society Series B}, 73(2):123-214, 2011.

\bibitem{chen2} X. Li, T.-K. L. Wong, R. T. Q. Chen, and D. K. Duvenaud. Scalable gradients for stochastic differential equations \textit{International Conference on Artificial Intelligence and Statistics}, 2020

\bibitem{valentin} V. Niess, A. Barnoud, C. Carloganu, and E. Le Menedeu. Backward Monte-Carlo applied to muon transport. \textit{Comput. Phys. Comm.}, 229(54), 2018.

\bibitem{pontrjagin} L. S. Pontryagin, E. F. Mishchenko, V. G. Boltyanskii, and R. V. Gamkrelidze. \textit{The mathematical theory of optimal processes}. John Wiley S., New York, London, 1962.

\bibitem{chris} C. Rackauckas, Y. Ma, J. Martensen, C. Warner, K. Zubov, R. Supekar, D. Skinner, and A. Ramadhan. Universal differential equations for scientific machine learning. \textit{preprint arXiv:2001.04385}, 2020.

\bibitem{tao} M. Tao. Explicit high-order symplectic integrators for charged particles in general electromagnetic fields. \textit{J. of Comp. Phys.}, 327:245-251, 2016.

\end{thebibliography}
\end{document}